\documentclass{aa}  

\usepackage{multirow}
\usepackage{graphicx}
\usepackage{txfonts}
\usepackage[pdftitle={Radial velocity constraints on the long-period transiting planet Kepler-1625\,b with CARMENES}, colorlinks = true, breaklinks = true, citecolor = blue, linkcolor = blue, urlcolor = blue, pdfauthor = {Anina Timmermann}]{hyperref}

\begin{document}

   \title{Radial velocity constraints on the long-period transiting planet Kepler-1625\,b with CARMENES}


   \author{Anina Timmermann\inst{1}
          \and
          Ren\'{e} Heller\inst{2}
          \and
          Ansgar Reiners\inst{1}
          \and
          Mathias Zechmeister\inst{1}
          }

   \institute{Institut f{\"u}r Astrophysik, Georg-August-Universit{\"a}t G{\"o}ttingen, Friedrich-Hund-Platz 1, 37077 G{\"o}ttingen, Germany\\
   \email{t.timmermann@stud.uni-goettingen.de}
         \and
Max Planck Institute for Solar System Research, Justus-von-Liebig-Weg 3, 37077 G\"ottingen, Germany
             }

   \date{Received 16/12/2019; Accepted 28/01/2020}

 
  \abstract
   {The star Kepler-1625 recently attracted considerable attention when an analysis of the stellar photometric time series from the Kepler mission was interpreted as showing evidence of a large exomoon around the transiting Jupiter-sized planet candidate Kepler-1625\,b. The mass of Kepler-1625\,b, however, has not been determined independently and its planetary nature has formally not been validated. Moreover, Kepler's long-period Jupiter-sized planet candidates, like Kepler-1625\,b with an orbital period of about 287\,d, are known to have a high false alarm probability. Hence, an independent confirmation of Kepler-1625\,b is particularly important.}
   {We aim to detect the radial velocity (RV) signal imposed by Kepler-1625\,b (and its putative moon) on the host star or, as the case may be, determine an upper limit on the mass of the transiting object (or the combined mass of the two objects).}
   {We took a total of 22 spectra of Kepler-1625 using CARMENES, 20 of which were useful. Observations were spread over a total of seven nights between October 2017 and October 2018, covering $125\,\%$ of one full orbit of Kepler-1625\,b. We used the automatic Spectral Radial Velocity Analyser (SERVAL) pipeline to deduce the stellar RVs and uncertainties. Then we fitted the RV curve model of a single planet on a Keplerian orbit to the observed RVs using a $\chi^2$ minimisation procedure.}
   {We derive upper limits on the mass of Kepler-1625\,b under the assumption of a single planet on a circular orbit. In this scenario, the $1\,\sigma$, $2\,\sigma$, and $3\,\sigma$ confidence upper limits for the mass of Kepler-1625\,b are $2.90\,M_{\rm J}$, $7.15\,M_{\rm J}$, and $11.60\,M_{\rm J}$, respectively ($M_{\rm J}$ being Jupiter's mass). An RV fit that includes the orbital eccentricity and orientation of periastron as free parameters also suggests a planetary mass but is statistically less robust.}
   {We present strong evidence for the planetary nature of Kepler-1625\,b, making it the 10th most long-period confirmed planet known today. Our data does not answer the question about a second, possibly more short-period planet that could be responsible for the observed transit timing variation of Kepler-1625\,b.
   }


   \maketitle
%
\section{Introduction}

The stellar system Kepler-1625 (KIC\,4760478, KOI\,5084) has become famous for its proposed candidate of an extrasolar moon around the transiting Jupiter-sized object Kepler-1625\,b \citep{TeacheyKEPLER,TeacheyHUBBLE}. If confirmed, this moon would be the first known exomoon. For now, however, the exomoon interpretation remains subject to debate \citep{RodenbeckEXOMOON,HellerHUBBLE,Kreidberg}.

The abundance of moons within the solar system suggests that there is also a plethora of moons around the thousands of exoplanets known today. These yet to be discovered exomoons are interesting objects, given their potential to allow insights into planet formation \citep{HellerEXOMOON} and possibly the spin orientation of their host planets \citep{MartinEXOMOON}. Moons have also been suggested as habitats beyond the solar system \citep{1997Natur.385..234W,2015A&A...578A..19H}, possibly sustained by the tidal heating driven by their host planets even far beyond the stellar habitable zone \citep{1987AdSpR...7..125R,2006ApJ...648.1196S,2013AsBio..13...18H,2014AsBio..14...50H} defined for planets \citep{1993Icar..101..108K}. The confirmation of the exomoon around Kepler-1625\,b would thus have implications for the field of exoplanet research as a whole and possibly even for astrobiology.

Here we want to take one step back from the exomoon scenario around Kepler-1625 and its Jupiter-sized transiting object and address the question of whether Kepler-1625\,b is actually a planet. \citet{2013ApJ...766...81F} found that the false positive rate of planet candidates as a function of planetary radius has a peak in the Jupiter-sized regime with a value of 17.7\,\% for planets with radii between 6 and 22 Earth radii. \citet{HellerEXOMOON} showed that the combined uncertainties in the radius measurements of the star and the planet propagate into the possibility of Kepler-1625\,b being rather a brown dwarf or possibly even  a very-low-mass star. That said, a preliminary Bayesian analysis of the combined transit photometry from the Kepler and Hubble space telescopes by \citet{2019arXiv190411896T} resulted in a posterior distribution of the planetary mass ($M_{\rm p}$) with a peak at $2.99_{-1.83}^{+2.86}\,M_{\rm J}$ ($M_{\rm J}$ being the mass of Jupiter) and a median value of $3.91\,M_{\rm J}$. The inference of the planetary mass was based on two methods: 1.) an empirical probabilistic mass-radius relation for the planet as implemented in the \textsc{forecaster} software \citep{2017ApJ...834...17C}; 2.) using \textsc{forecaster} for the moon and computing the planetary mass via the moon-to-planet mass ratio as fitted with photodynamical modelling \citep{TeacheyKEPLER}. Assuming a stellar mass ($M_\star$) of $1.079_{-0.138}^{+0.100}\,M_\odot$ \citep{2017ApJS..229...30M}, where $M_\odot$ is the solar mass, the expected radial velocity (RV) amplitude of a $3\,M_{\rm J}$ planet on a 287\,d circular orbit \citep{TeacheyKEPLER} is about $88\,{\rm m\,s}^{-1}$.

This signal could be in reach of the ``Calar Alto high-Resolution search for M dwarfs with Exoearths with Near-infrared and optical {\'E}chelle Spectrographs'' (CARMENES) at the 3.5m telescope at Calar Alto Observatory \citep{2018SPIE10702E..0WQ,2018A&A...612A..49R}. In fact, CARMENES has recently reached the $1\,{\rm m\,s}^{-1}$ precision level that resulted in the detection of two Earth-mass planets around Teegarden's star \citep{2019A&A...627A..49Z}. Teegarden's star, however, is a relatively bright M dwarf with visual and near-infrared magnitudes of $V~=~15.08\,(\pm0.12)$ (value from \citealt{2015AAS...22533616H} as cited by \citealt{2019A&A...627A..49Z}) and $J~=~8.39\,(\pm 0.03)$ \citep{2003yCat.2246....0C}, whereas Kepler-1625 is a slightly evolved ($R_\star~=~1.793_{-0.488}^{+0.263}\,R_\odot$), significantly fainter solar type star with a Gaia magnitude of $G~=~15.7627\,(\pm\,0.0005)$ \citep{2018yCat.1345....0G} and $J~=~14.364\,(\pm\,0.032)$ \citep{2003yCat.2246....0C}. Prior to our observations, the stellar properties of Kepler-1625 suggested an RV precision of approximately $60\,{\rm m\,s}^{-1}$ in the wavelength range from 650\,nm to 750\,nm, based on the photon limit alone \citep{ReinersZechmeister}.\footnote{Computed for an effective temperature of $T_{\rm eff}~=~5600$\,K, $J~=~14$, a telescope aperture of $\diameter~=~3.5$\,m, a signal-to-noise ratio of S/N~=~4, a resolution of $R~=~~90,000$, and an exposure time $t_{\rm obs}~=~20$\,min using the online Radial Velocity Precision Calculator at \href{http://www.astro.physik.uni-goettingen.de/research/rvprecision/}{www.astro.physik.uni-goettingen.de/research/rvprecision}.}

\section{Methods}

\subsection{Observations}

CARMENES provides two sets of spectra from its two {\'E}chelle spectrographs, or ``channels''. One channel covers the near infrared ($0.96\,\mu{\rm m}\,\leq\,\lambda\,\leq\,1.71\,\mu{\rm m}$, $\lambda$ being the wavelength), the other one covers the visible light \citep[$0.52\,\mu{\rm m}\,\leq\,\lambda\,\leq\,0.96\,\rm \mu m$;][]{CARMENES}. Data from both channels were processed using the CARACAL pipeline (of M.~Zechmeister) by performing a dark/bias correction, order tracing, flat-relative optimal extraction, cosmic ray correction, and wavelength calibration.

We took a total of 22 spectra of Kepler-1625 with CARMENES. Table~\ref{tab:CARMobservations} lists the observation dates of the exposures in local time and Barycentric Julian Date (BJD), the mean S/N over all spectral orders, the non-calibrated RVs with their corresponding uncertainties, and the observation ID of every exposure. Observations took place during seven nights between 25.10.2017 and 23.10.2018 (CAHA proposal PI R.~Heller). The time spanned by our observations covers about 125\,\% of one orbit of Kepler-1625\,b. Three of the observations were scheduled to be within one week of an expected planetary transit. We consider a model of a single planet on a circular orbit, where the RV during planetary transits is known to be $0\,{\rm m\,s}^{-1}$. This allows us to calibrate our RV measurements. Two observations were close to the expected peak RV values and two at intermediate orbital phases.

One spectrum, taken on 30.06.2019, suffered from bad weather conditions that resulted in a low S/N, which is why we did not use it for our analysis. This spectrum did not receive an ID and is labelled with an asterisk in Table~\ref{tab:CARMobservations}. For data quality reasons, we also did not take the spectrum with ID 9 into account for our RV analysis (see Sect.~\ref{sec:RV}).

Each observing night resulted in three spectra with an exposure of 20\,min each. This setup was chosen to reduce the effects of potential stellar oscillation and granulation on the spectrum, to facilitate the detection of cosmic particle hits, and to allow for the partial rejection of data that could be affected by clouds or other weather effects.

\begin{figure}
\centering
\includegraphics[width=1\linewidth]{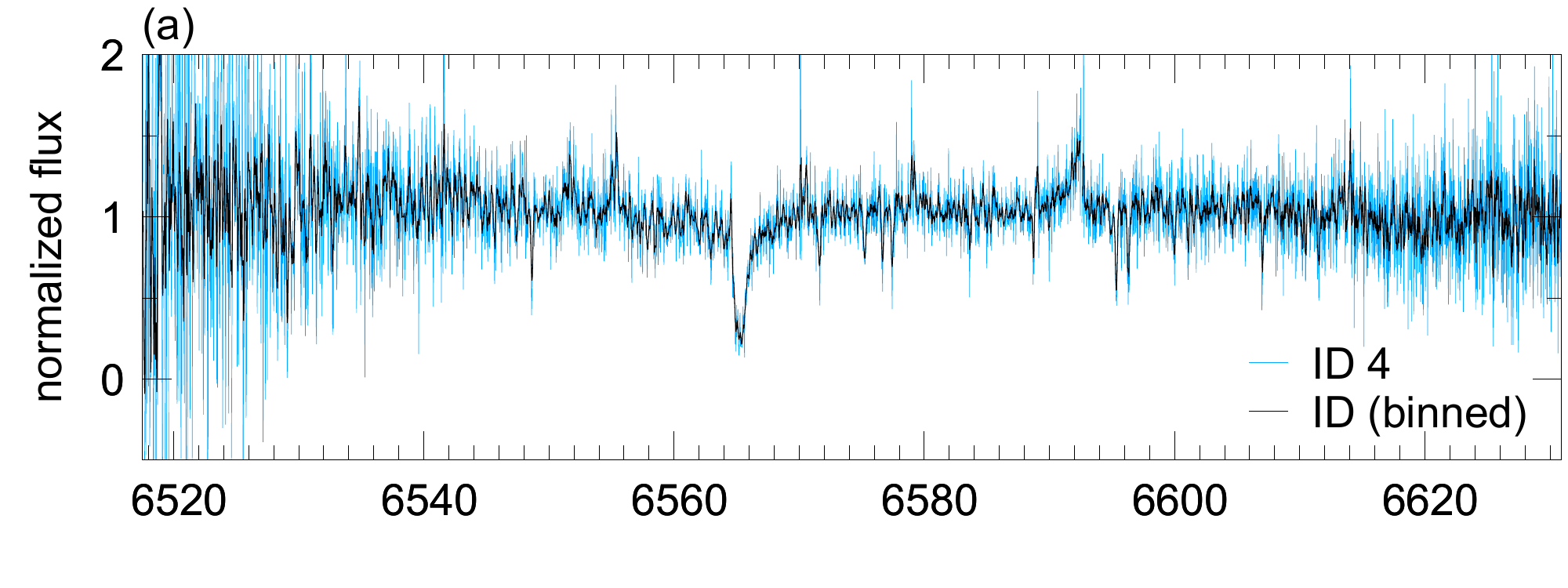}\\
\includegraphics[width=1\linewidth]{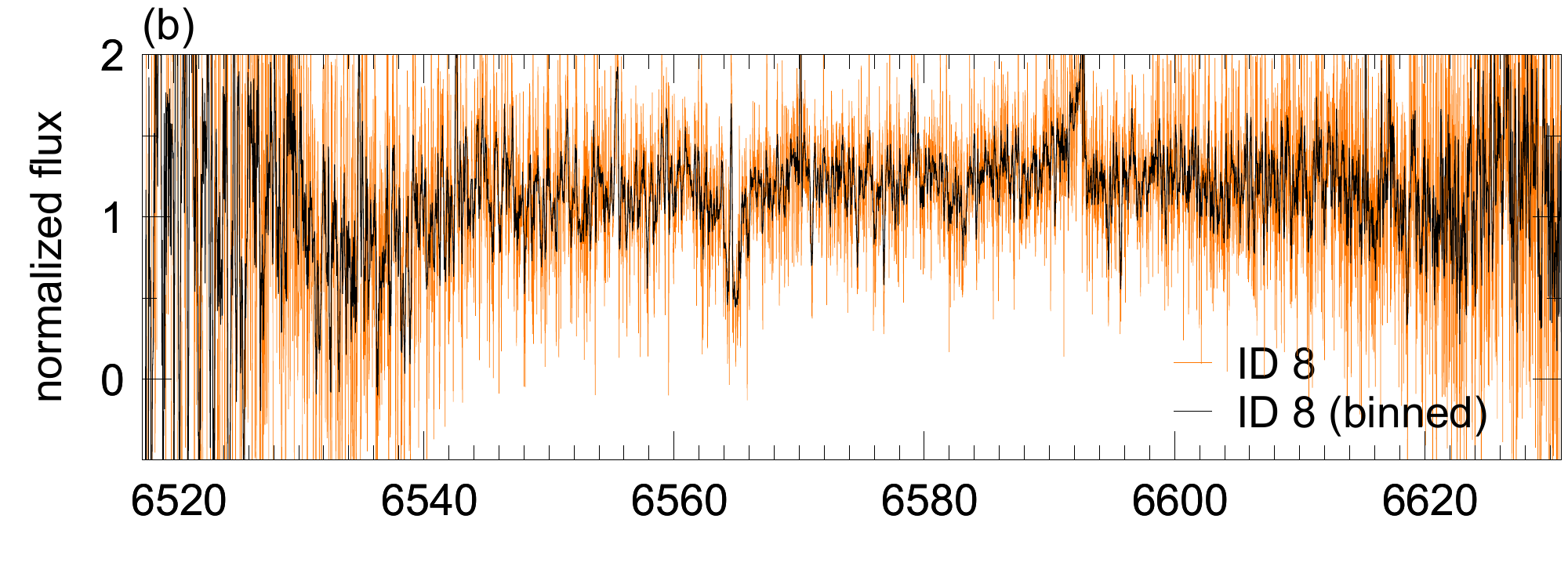}\\
\includegraphics[width=1\linewidth]{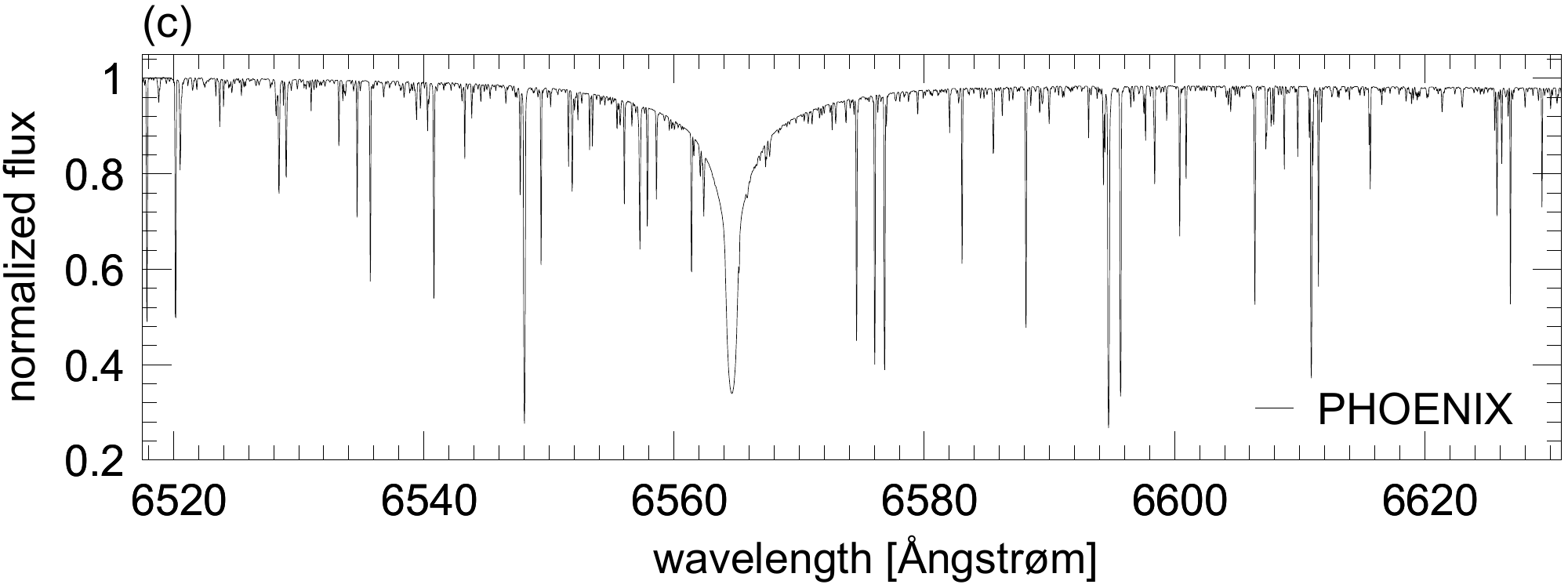}
\caption{Sample CARMENES spectra (order 25) of Kepler-1625 and comparison to synthetic spectrum. \textbf{(a)} CARMENES spectrum of a relatively high-S/N observation (ID 4) with blue showing the data from the CARACAL pipeline. The black line represents a binning of the spectrum for illustrative purpose only and to assist the human eye in separating spectral lines from noise. We did not use the binned data for our spectral analysis. Seven data points were used for binning. \textbf{(b)} CARMENES data of a relatively low-S/N observation (ID 8) with orange showing the CARACAL pipeline data and black showing the binned data. \textbf{(c)} Synthetic PHOENIX spectrum of a model star akin to Kepler-1625, with $T_{\rm eff}=5600$, ${\log}(g)=4.0$, and ${\rm [Fe/H]}=0$.}
\label{fig:spec_ord25}
\end{figure}

\begin{figure*}
\centering
\includegraphics[angle= 0,width=.67\linewidth]{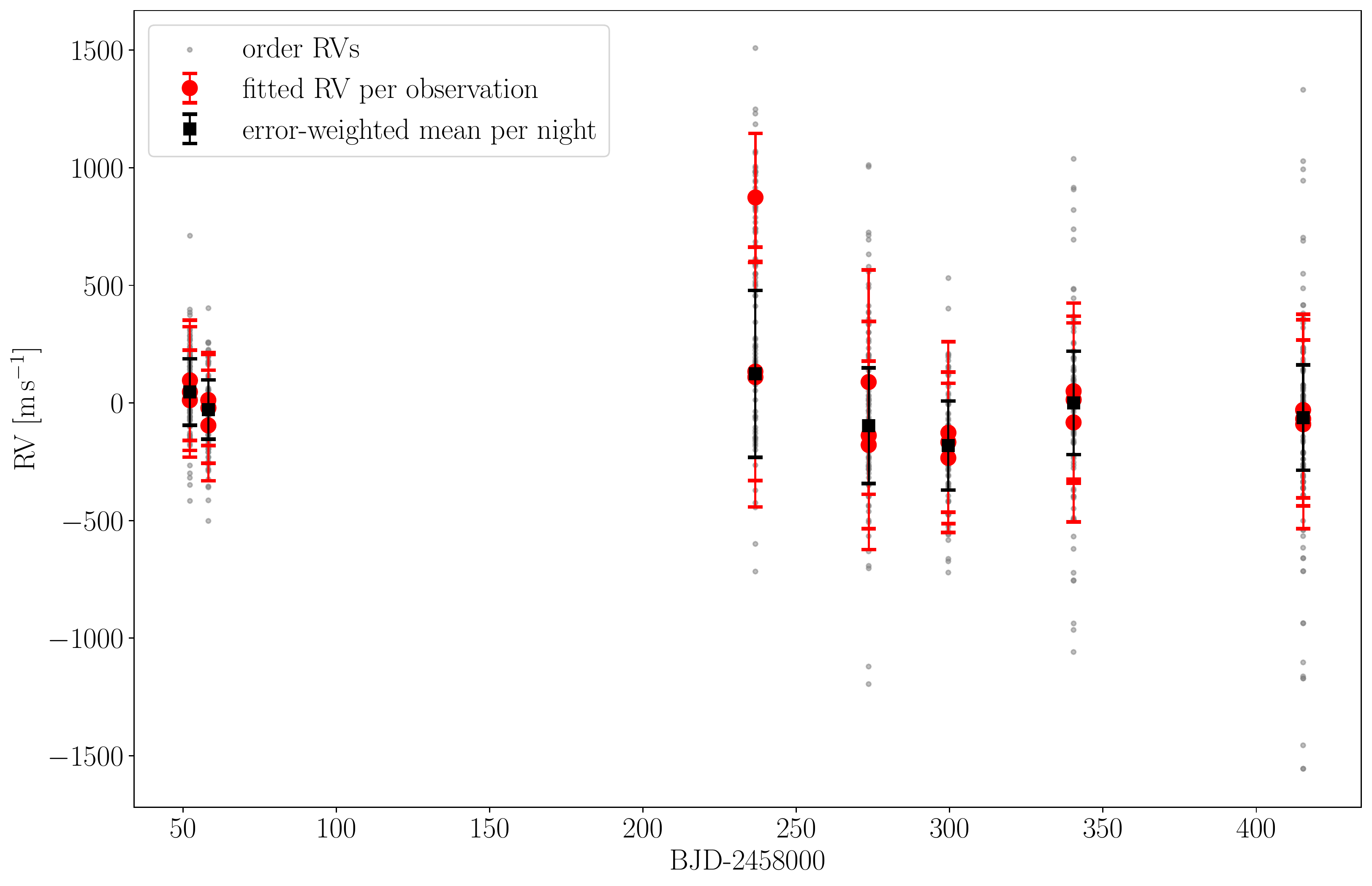}
\caption{RV measurements of Kepler-1625. Grey dots represent the measurements derived with the SERVAL pipeline for spectral orders 20 to 48 of the visible channel of CARMENES. Three red circles for each observation night show the peak values of the respective normal distribution after fitting to the co-added normal distributions of all orders. Black squares with error bars symbolise the nightly error-weighted mean RV values and mean errors.}
\label{fig:orders}
\end{figure*}


Figure~\ref{fig:spec_ord25} displays two sample spectra of the spectral order 25 around the prominent H$\alpha$ line, with panel (a) showing the spectrum with the highest S/N (ID 4) and panel (b) illustrating a rather low-S/N observation (ID 8). Panel (c) shows a model spectrum for a Kepler-1625-like star from the PHOENIX spectral library \citep{2013A&A...553A...6H}. Our near-infrared observations had very low S/N, so we decided to work with the spectra from the visible channel only. The spectra were structured in a total of 61 overlapping {\'E}chelle spectral orders. We found that spectral orders 20 to 48 had sufficiently high S/N values to deliver useful spectral information. The remaining orders typically had $\rm S/N <2$ and their RVs differed significantly and systematically from the error-weighted mean RV. Due to the apparent faintness of Kepler-1625, the order-averaged S/N for the 20 useful spectra were relatively low, ranging between 2.8 and 6.9 with an average value of 4.1.

\begin{table*}[t]
\centering
 \def\arraystretch{1.05}
  \caption{Logbook of our CARMENES observations of Kepler-1625.}
  \label{tab:CARMobservations}
 \begin{tabular}{c|c|c|c|c|c} 
Date & BJD & S/N & RV [${\rm m\,s}^{-1}$]$^\dag$ & Uncertainty [${\rm m\,s}^{-1}$] &  ID \\ \hline \hline
\multirow{3}{*}{25.10.17} & 2458052.26775 & 5.37 & -48.80 & 213.28 & 1 \\ \cline{2-6}
 & 2458052.28352 & 4.27 & -13.51 & 277.22 & 2 \\ \cline{2-6}
 & 2458052.29859 & 5.14 & 35.56 & 255.88 & 3 \\ \hline
\multirow{3}{*}{31.10.17} & 2458058.28172 & 6.92 & -48.15 & 193.20 & 4 \\ \cline{2-6}
 & 2458058.30206 & 4.88 & -156.14 & 235.08 & 5 \\ \cline{2-6}
 & 2458058.31763 & 5.37 & -81.55 & 235.59 & 6 \\ \hline
 \multirow{3}{*}{28.04.18} & 2458236.66051 & 3.53 & 72.84 & 463.50 & 7 \\ \cline{2-6}
 & 2458236.67569 & 3.40 & 49.30 & 552.19 & 8 \\ \cline{2-6}
 & 2458236.69211 & 7.06 & 813.10 & 272.56 & 9 \\ \hline
 \multirow{3}{*}{04.06.18} & 2458273.60239 & 3.49 & 28.41 & 477.13 & 10 \\ \cline{2-6}
 & 2458273.61722 & 3.59 & -239.04 & 356.24 & 11 \\ \cline{2-6}
 & 2458273.63306 & 3.51 & -198.67 & 485.05 & 12 \\ \hline
 \multirow{4}{*}{30.06.18} & 2458299.54861 & 0.83 & * & * & * \\ \cline{2-6}
 & 2458299.57736 & 3.88 & -294.20 & 317.47 & 13 \\ \cline{2-6}
 & 2458299.59385 & 4.32 & -227.46 & 297.66 & 14 \\ \cline{2-6}
 & 2458299.60959 & 3.92 & -187.11 & 386.82 & 15 \\ \hline
 \multirow{3}{*}{09.08.18} & 2458340.42432 & 3.24 & -143.30 & 423.12 & 16 \\ \cline{2-6}
 & 2458340.43902 & 4.27 & -10.43 & 374.32 & 17 \\ \cline{2-6}
 & 2458340.45349 & 4.37 & -46.58 & 354.83 & 18 \\ \hline
 \multirow{3}{*}{23.10.18} & 2458415.27194 & 2.91 & -90.50 & 407.24 & 19 \\ \cline{2-6}
 & 2458415.29215 & 2.92 & -128.29 & 335.98 & 20 \\ \cline{2-6}
 & 2458415.30818 & 2.80 & -150.87 & 443.99 & 21 \\ \hline
\end{tabular}\\
\tablefoot{\dag \ Values not corrected for the system's intrinsic RV offset of $-60.43\,{\rm m\,s}^{-1}$ with respect to the solar system barycenter. * Spectrum not analysed.}
\end{table*}

\subsection{Radial velocity analysis}
\label{sec:RV}

We analysed the data with the Spectral Radial Velocity Analyser \citep[SERVAL;][]{ZechmeisterSERVAL} with the aim of deriving the stellar RVs in each spectral order of every exposure. SERVAL derives RVs by comparing the data of individual spectral orders to a high S/N template of the observed star, which is created by co-adding B-spline regressions of all available data sets \citep{ZechmeisterSERVAL}. The code was run for the spectral orders 20 to 48, with an oversampling factor for the creation of the template of 0.3. The oversampling factor corresponds to the number of B-spline knots per data point, which we adjusted manually to retain as much spectral information as possible while simultaneously reducing the effect of noise.

We found that the default SERVAL scheme to weight the RV information from different orders produced unrealistic results for our data because of the low S/N. As a consequence, we decided to model the RV of each spectral order as a normal distribution centred on the SERVAL RV estimate of a given order with the SERVAL error estimate as the standard deviation. The resulting normal distributions of spectral orders 20 to 48 were co-added and normalised for each exposure. In this co-adding process, all orders were given the same weight in the sense that all the normal distributions from each order were normalised to an area of one. In the next step, we fitted a normal distribution function to the co-added RV distribution. We took the peak position of the fitted curve as the reference RV value and its standard deviation as the uncertainty of a given exposure. Finally, the error-weighted mean and mean standard deviations of the exposures of one night were taken as the RV value and corresponding uncertainty of that night.

In this process, the spectrum with ID 9 turned out to have suffered from unknown systematic effects, which resulted in a suspiciously high average S/N over all orders, a large S/N dispersion between the orders, and an extremely high RV value of about $800\,{\rm m\,s}^{-1}$. For comparison, the RV values that we derived from remaining spectra differ by $\lesssim\,250\,{\rm m\,s}^{-1}$. We also noticed a sinusoidal variation of the RVs as a function of the order number, which was not present for any other spectrum and which cannot be explained by the astrophysical processes that we are interested in. Hence, we consider this spectrum with ID 9 as compromised and rejected it for our further RV study.

We considered a Keplerian model of a single planet on a circular orbit with the planetary mass ($M_{\rm p}$) as the single free parameter. In this model, the stellar RV during the planetary transit is known to be zero once corrected for the system's intrinsic RV with respect to the solar system. Accordingly, we used the mean RV values of the three observations closest to a planetary transit (IDs 1-3, 4-6, and 16-18) for a zero-point calibration of the stellar RV by means of a linear regression. We obtained an offset of $-60.43\,{\rm m\,s}^{-1}$. Then we fitted the model to the calibrated RV values. The transit times, orbital period ($P~=~287.3776\,\pm\,0.0024$\,d)\footnote{This value was derived using Markov chain Monte Carlo simulations of the a total of four transits of Kepler-1625\,b observed with the Kepler and Hubble space telescopes \citep{HellerHUBBLE}. It is based on the assumption of a planet-only model without any perturbations from the hypothesised exomoon.}, and stellar mass $M_{\star}~=~1.079_{-0.138}^{+0.100}\,M_\odot$ \citep{2017ApJS..229...30M} were known with sufficiently high precision and thus fixed at their nominal values. We established upper mass on $M_{\rm p}$ using a $\chi^2$ minimisation and a statistical analysis of the $\chi^2$ distribution, based on the $\Delta \chi^2$-values for different confidence levels.


\section{Results}

In Fig.~\ref{fig:orders} we show the RV values of Kepler-1625 derived with SERVAL that we corrected for the zero point of the RVs. Small grey dots represent the RV estimates of each spectral order, with the respective error bars being omitted for the sake of clarity. The scatter of the orders is several $100\,{\rm m\,s}^{-1}$ every night, indicative of the low S/N of the data. That said, this first insight into the data suggests that our combination of the information from all orders (Sect.~\ref{sec:RV}), which naturally increases our sensitivity, could reveal the RV signal of a planet with a mass near the planet-brown dwarf boundary. A $13\,M_{\rm J}$ transiting planet in a 287\,d orbit around Kepler-1625\,b would cause an RV amplitude of about $380\,{\rm m\,s}^{-1}$.

The large red circles in Fig.~\ref{fig:orders}, three for every night with observations, illustrate the mean value of orders 20-48 of each 20\,min exposure and the resulting standard deviation as derived with our co-adding and fitting of normal functions. Note  the large value of approximately $800\,{\rm m\,s}^{-1}$ near 237\,d (BJD-2,458,000), which corresponds to the spectrum with ID 9 that we did not consider for our derivation of the RV signal. The black squares represent the error weighted mean value of each night.

We computed the $\chi^2$ distribution as a function of $M_{\rm p}$ for our fitting procedure of a one-planet Keplerian model with a circular orbit. The $\chi^2$ minimum value of $1.43$ is found at $M_{\rm p}=0$. The confidence levels of $1\sigma$, $2\sigma$, and $3\sigma$ for upper mass limits of $2.90\,M_{\rm J}$, $7.15\,M_{\rm J}$, and $11.60\,M_{\rm J}$, respectively, are indicated with dashed lines. In Fig.~\ref{fig:RVcurve} we visualise the corresponding RV curves. Note that the RV values near the expected transit times of Kepler-1625\,b at about 52\,d, 58\,d, and 340\,d (BJD-2,458,000), the latter of which were fixed in our $\chi^2$ minimisation procedure, agree within about $75\,{\rm m\,s}^{-1}$, well within their error bars.

In summary, the data can best be explained in the absence of any RV signal induced by Kepler-1625\,b and our upper limits on the planetary mass confirm that Kepler-1625\,b must be a planet under the assumption of a single planet around Kepler-1625.

\section{Discussion}

\begin{figure}
\centering
\includegraphics[angle= 0,width=1.01\linewidth]{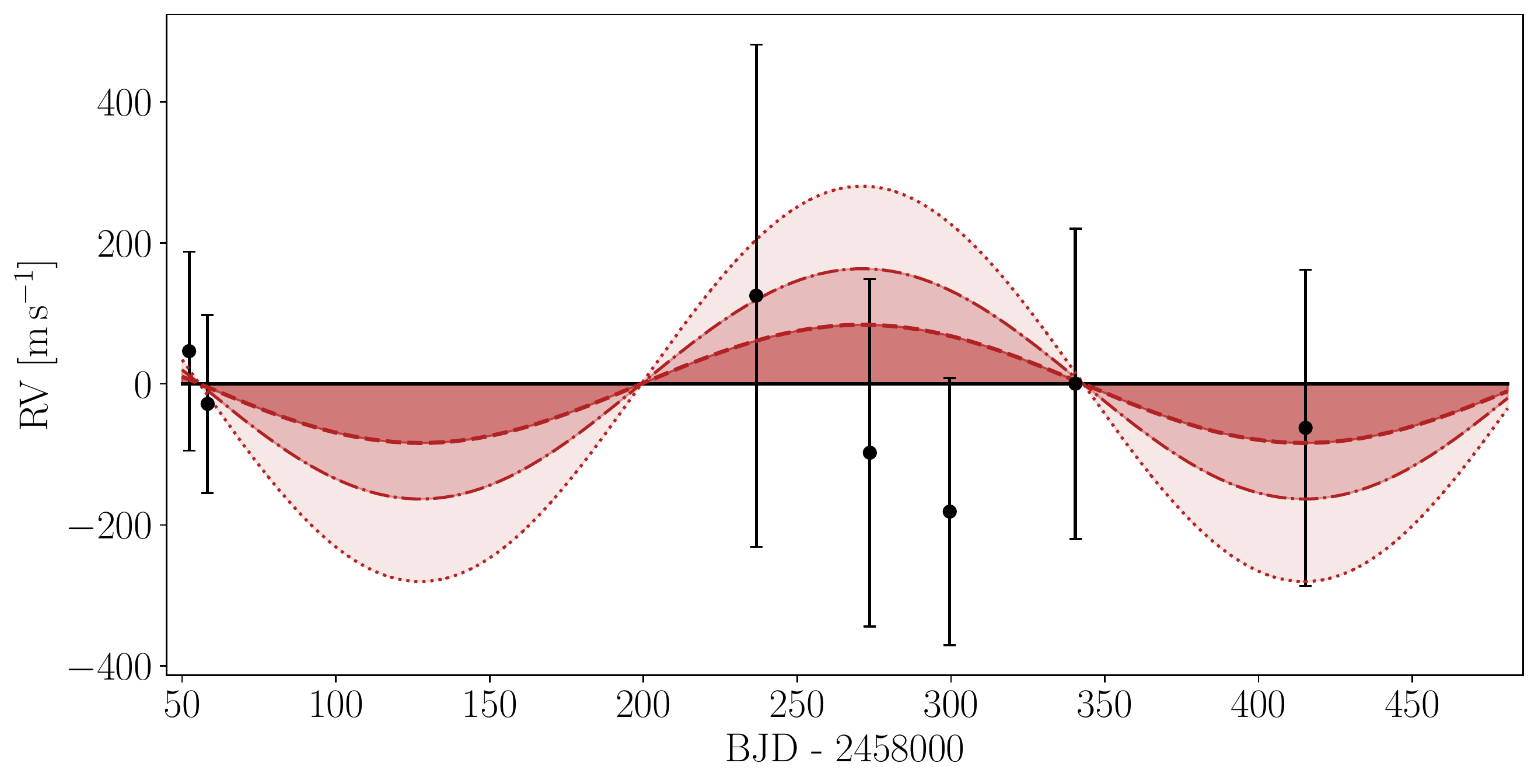}
\caption{Derived RV values and RV curve for the found planetary masses (2.90, 7.15, 11.60\,$M_{\rm J}$) of different confidence levels (1$\sigma$, $2\sigma$, $3\sigma$). The lighter the shade of the colour, the higher the confidence regarding the corresponding mass limit.}
\label{fig:RVcurve}
\end{figure}

The computed $\chi^2$ values are relatively small, given the number of five free parameters. This is due to the relatively large formal error bars compared to the intrinsic scatter of the data (see Fig.~\ref{fig:RVcurve}). This observation suggests that the formal error bars that we derived are overestimated and that our resulting upper limits for $M_{\rm p}$ can be regarded as conservative.

For the purpose of completeness, we also investigated non-circular orbits, in which the orbital eccentricity ($e$) and the argument of the periastron ($\omega$) were free fitting parameters with flat priors. As a result, we find a mass of Kepler-1625\,b that is still in the planetary regime but the formal error bars in each of the fitting parameters are so large that an eccentric orbit is effectively unconstrained. The reason is in the small number of RV measurements, which is comparable to the number of fitting parameters. As a consequence, we do not consider these results as statistically robust and thus refrain from a more detailed analysis. Moreover, using priors from the transit fit would result in marginal changes of the posterior distributions. These small changes in the formal best fit values are probably smaller than the systematic errors in our method, which is why we discard a more in-depth analysis.

The mere non-detection of a RV signal could still allow the observed transit signal to be due to an astrophysical false positive if these scenarios cannot be ruled out otherwise. As shown by \citet{2016ApJ...822...86M}, however, the false positive probability of Kepler-1625\,b of being caused by either an unblended eclipsing binary, or a hierarchical eclipsing binary, or a background/foreground eclipsing binary is $8.5~{\times}~10^{-3}\,(\pm\,5.1~{\times}~10^{-4})$. Moreover, the probability of the signal emerging from the target star is 1 \citep{2016ApJ...822...86M} and the transit sequence has been successfully modelled as resulting from a Jupiter-sized object around Kepler-1625, be it with a moon or without a moon \citep{TeacheyKEPLER,RodenbeckEXOMOON,TeacheyHUBBLE,HellerHUBBLE}. The only remaining possibility is a Jupiter-sized transiting object around Kepler-1625, which we here constrain to have a mass in the planetary regime.

\section{Conclusions}
We have analysed a total of 21 CARMENES spectra of the star Kepler-1625 distributed over seven nights in a time span of approximately one year, or 125\,\% of the orbital period of Kepler-1625\,b. Our examination of the RVs and their error bars, combined with the previous rejection of an astrophysical false positive scenario, allows us to confirm the planetary nature of the transiting Jupiter-sized object Kepler-1625\,b. Under the assumption of a single planet on a circular orbit, its mass is lower than $2.90\,M_{\rm J}$, $7.15\,M_{\rm J}$, or $11.60\,M_{\rm J}$ with a confidence of $1\sigma$, $2\sigma$, or $3\sigma$, respectively. We have also investigated eccentric orbits suggestive of a planetary mass, but this fit did not reduce the $\chi^2$ value substantially and we consider it less robust given the small amount of measurements.

Our results make Kepler-1625\,b one of the most long-period transiting planets ever detected and confirmed via the RV method. This result would also hold if the planet were orbited by a Neptune-mass moon. The mass of such a moon would be of the order of a few percent of the planet's mass, assuming a Neptune-like mass. This mass of the satellite would need to be subtracted from the total mass estimates that we provide, thereby decreasing the actual mass of the planet by a few percent. We cannot, however, exclude a second massive planet in the system, which has been hypothesised to explain the observed transit timing variations of Kepler-1625\,b \citep{TeacheyHUBBLE,HellerHUBBLE}. This hypothesis could best be explored with new and more high-quality (S/N$~{\gtrsim}~20$) RV observations that are particularly sensitive to short-period planets, e.g. taken during successive nights over the course of several weeks. If successful in the hunt for a second, Jupiter-mass non-transiting planet \citep[as proposed by][]{HellerHUBBLE}, then these observations would have the potential to reject the exomoon hypothesis for Kepler-1625\,b by explaining the observed transit timing variations.

\begin{acknowledgements}
The authors thank Tim-Oliver Husser for a helpful discussion about the spectral classification of Kepler-1625.

RH is supported by the German space agency (Deutsches Zentrum f\"ur Luft- und Raumfahrt) under PLATO Data Center grant 50OO1501.

CARMENES is an instrument for the Centro Astron\'omico Hispano-Alem\'an de
  Calar Alto (CAHA, Almer\'{\i}a, Spain). 
  CARMENES is funded by the German Max-Planck-Gesellschaft (MPG), 
  the Spanish Consejo Superior de Investigaciones Cient\'{\i}ficas (CSIC),
  the European Union through FEDER/ERF FICTS-2011-02 funds, 
  and the members of the CARMENES Consortium 
  (Max-Planck-Institut f\"ur Astronomie,
  Instituto de Astrof\'{\i}sica de Andaluc\'{\i}a,
  Landessternwarte K\"onigstuhl,
  Institut de Ci\`encies de l'Espai,
  Institut f\"ur Astrophysik G\"ottingen,
  Universidad Complutense de Madrid,
  Th\"uringer Landessternwarte Tautenburg,
  Instituto de Astrof\'{\i}sica de Canarias,
  Hamburger Sternwarte,
  Centro de Astrobiolog\'{\i}a and
  Centro Astron\'omico Hispano-Alem\'an), 
  with additional contributions by the Spanish Ministry of Economy, 
  the German Science Foundation through the Major Research Instrumentation 
    Programme and DFG Research Unit FOR2544 ``Blue Planets around Red Stars'', 
  the Klaus Tschira Stiftung, 
  the states of Baden-W\"urttemberg and Niedersachsen, 
  and by the Junta de Andaluc\'{\i}a.
  Based on data from the CARMENES data archive at CAB (INTA-CSIC).
 
\end{acknowledgements}

%
%

\bibliographystyle{aa}
\bibliography{references}

\end{document}